\documentclass[showpacs,preprintnumbers,amsmath,amssymb]{revtex4}
\usepackage{graphicx}
\usepackage{dcolumn}
\usepackage{bm}
\usepackage{tensor}
\begin{document}
\title{Superfluid rotation sensor with helical laser trap.}
\author{A.Yu.Okulov} 
\email{alexey.okulov@gmail.com}
\homepage{http://okulov-official.narod.ru}
\affiliation{Russian Academy of Sciences, 
119991, Moscow, Russia} 
 
\date{\ November 12, 2012}
 
\begin{abstract}
{ The macroscopic quantum states of the dilute bosonic ensemble 
in helical laser trap at the temperatures about 
$10^{-6}\bf {K}$ are considered 
in the framework of the Gross-Pitaevskii equation. 
The helical interference pattern is composed of the two 
counter propagating Laguerre-Gaussian optical vortices with opposite 
orbital angular momenta $\ell \hbar$ and this pattern is  
driven in rotation via angular Doppler effect. 
Macroscopic observables including linear momentum and 
angular momentum 
of the atomic cloud are evaluated explicitly. 
It is shown that rotation of reference frame is transformed 
into translational motion of the twisted matter wave. 
The speed  
of translation equals the group velocity of twisted 
wavetrain $V_z= \Omega\ell/ k$ and alternates with a sign of the 
frame angular velocity $\Omega$ and helical pattern handedness $\ell$. 
We address detection of this effect using currently 
accessible laboratory equipment with emphasis on 
the difference between 
quantum and classical fluids.}
\end{abstract}


\maketitle

\section{Introduction}

The superfluids are often considered as an absolute 
reference frame. This means that when a density of condensate 
fraction is large enough $\rho_s \rightarrow \rho $ 
the quantum fluid 
confined in annular container (fig.1a) and cooled below phase 
transition temperature $T_{\lambda}$ 
proves to be in irrotational state 
with zero angular momentum $L_z=0$ regardless 
to the changes of container's angular velocity 
$\Omega $\cite {Fairbank:1967}. 
This happens at rather slow rotation rates 
($\Omega /2\pi \le \cdot 10^{-3} Hz$). 
At a higher angular velocities a metastable 
persistent flow with nonzero angular momentum 
per particle $L_z = \ell \hbar$ is observable 
\cite {Feynman:1972}. The same holds for the quantum gas in 
multiply connected optical 
and magnetic traps \cite {Phillips:2011, Wright:2001}, 
where order parameter ${\Psi}$ is 
given by the Gross-Pitaevskii equation 
(GPE) \cite {Pitaevskii:1999}: 
\begin{equation} 
\label{GPE_rot_frame}
{i \hbar}{\frac {\partial {\Psi}}
{\partial t}} = -{\frac {\hbar^2}{2 m}} 
\Delta {\Psi} +{  U_{ext}{(z,r,\theta,t)}}{\:}{\Psi} 
+{g}|{\Psi}|{\:}^2 {\Psi},
\end{equation}
where $m$ is atomic mass, $g=4 \pi \hbar^2 a_S/m$ is 
interaction constant, $a_S$ is S-wave scattering length, 
$U_{ext} $ is external confining potential, ${(z,r,\theta)}$ 
are cylindrical coordinates. For the quantum 
liquid confined by an annular container (fig.1a) with 
impenetrable walls the potential 
is a step-like function: 
\begin{equation}
\label{confining_container}
\ { U_{an}{(z,r,\theta)}} \sim U_{0_{an}} \cdot
\Theta (|r-D|-(d/2)) \cdot \Theta (|z|-L_c/2),{\:}{\:}
\end{equation}
where $\Theta$ is Heavyside step-function,
$D$ is radius of container, $d$ is a thickness of walls, $L_c$ 
is a length of container in $z$-direction, the potential 
wall $U_{0_{an}}\rightarrow \infty$ is assumed to be 
much higher then chemical potential $\mu$. For trapped 
quantum gases the similar multiply 
connected potential configurations with 
a smooth penetrable walls are used. For example the 
toroidal trap inside a tightly 
focused ($F/D_0 \cong 1$, fig.1b) Laguerre-Gaussian (LG) 
beam  forms the following trapping 
potential $U_{tor}$ in LG beam waist \cite {Phillips:2011, Wright:2001}: 
\begin{equation}
\label{TRAPOP_atomic_tor}
\ { U_{tor}{(z,r,\theta)}} \sim {\frac 
{U_0 \cdot r^{2|\ell|}}{(1+z^2/(k^2 {D_0}^4))} }
\exp \bigg{[}-{\frac {2r^2}{D_0^2(1+{ {z^2}/{k^2 D_0^4}})}}\bigg{]}
,{\:}{\:}
\end{equation} 
where $D_0$ is radius of LG beam, $k=2\pi/\lambda$, $\lambda$ 
is the optical trapping wavelength, $\ell$ is azimuthal quantum 
number (topological charge) of LG, $F$ is the focal length of 
objective, $U_0 \cong k_B T, T \sim 10^{-(3-6)} K$. 
This elongated torus has elliptical cross-section 
with $D_0^2/ \lambda$ (Rayleigh range)
width in $z$-direction and $\sim D_0$ width in $r$ (radial) direction. 
The current activity in 
toroidal traps is targeted to study Josephson phase 
dynamics and SQUID-like applications \cite{Phillips:2011}. 
  
The periodic sequence of toroidal traps was also implemented 
as stationary interference pattern of counter-propagating LG beams (fig.1c): 
\begin{equation}
\label{TRAPOP_atomic_tor_periodic}
\ { U_{tor_P}{(z,r,\theta)}} \sim {\frac 
{U_0 \cdot r^{2|\ell|}}{(1+z^2/(k^2 {D_0}^4))} }
\exp \bigg{[}-{\frac {2r^2}{D_0^2(1+{ {z^2}/{k^2 D_0^4}})}}\bigg{]}
[1+\cos(2kz)],{\:}{\:}
\end{equation}
where phase matching had been achieved by confocal parabolic mirrors 
\cite {Rempe:2007}. This sequence of the 
$\lambda/2$-separated toruses also has elliptical cross-section 
with $\lambda/2$ 
width in $z$-direction and $\sim D_0$ width in radial direction
\cite {Okulov:2008}. 

In rotating toroidal containers (fig.1a,b,c) the circular flows 
of the liquid $^4He $ 
\cite {Fairbank:1967,Feynman:1972} or 
degenerate quantum gas of a cold bosons in a toroidal 
trap \cite {Fetter:2009,Leggett:2001,Konotop:2007} have a quantized circulation 
of the velocity $\vec v(\vec r)$ along a closed contour $\Gamma$ embedded in a 
superfluid ensemble: 
\begin{equation}
\label{circulation}
\oint _{\Gamma} \vec v(\vec r) \cdot d \vec r = 
{\frac {\hbar}{m}}\oint _{\Gamma} 
\nabla \phi  \cdot d \vec r = n \kappa,{\:} 
\kappa={\frac {2\pi \hbar}{m}} {\:},
\end{equation}
where $\kappa$ is a quantum of circulation, $n$ is integer, 
$\phi$ is the argument 
of the macroscopic wavefunction $\Psi$ \cite {Pitaevskii:1999}. 
When reference frame rotates with container with angular velocity 
$\Omega$ the circulation acquires the additional Sagnac 
term \cite {Scully:1997}:
\begin{equation}
\label{circulation_in_rotframe}
\oint _{\Gamma} \vec v(\vec r) \cdot d \vec r = 
\oint _{\Gamma} {[\vec v(\vec r) - \vec {\Omega} \times \vec r ]}
\cdot d \vec r  
= n \kappa - 2 \vec {\Omega}\cdot \vec S,{\:} 
\end{equation}
where $\vec S$ is area enclosed by contour $\Gamma$. Hence 
the phase of wavefunction is shifted by 
$\delta \phi_{SF}=(m / \hbar)2 \Omega \cdot \vec S$. 
Taking into account de Broglie wavelength 
$\lambda_B=2 \pi \hbar / (m |\vec v|)$ of massive particle moving 
with speed $|\vec v|$  
the similarity between Sagnac shifts for massless photons 
$\delta \phi_{phot}$ \cite {Scully:1997,Okulov:2010josa} and matter 
waves becomes evident.The numerical comparison shows that of matter wave 
Sagnac interferometer has better accuracy: 
\begin{equation}
\label{Sagnac_comparison}
\delta \phi_{SF}=
{\frac {4 \pi \vec {\Omega} \cdot \vec S}{\lambda_B |\vec v|}},{\:} 
\delta \phi_{phot}=
{\frac {4 \pi \vec {\Omega} \cdot \vec S_{phot}}{\lambda c}}, 
{\frac {\delta \phi_{SF}}{\delta \phi_{phot}}}=
{\frac {\lambda m c {|\vec S|}}{h {|\vec S_{phot}|}}}
= {\frac {m c^2 {|\vec S|}}{h\nu {|\vec S_{phot}|}} }
\sim 10^{10} {\frac { {|\vec S|}}{ {|\vec S_{phot}|}} }, 
\end{equation}
because of smallness of photon's energy 
$h \nu=h \lambda/c \sim 1 eV$ compared 
to the rest mass of a typical atom. This advantage 
of the matter waves is a somewhat diminished by a substantially 
larger area $|\vec S_{phot}|$ enclosed by optical fibers compared 
to atomic waveguides $|\vec S|$  \cite {Scully:1997}. 
This motivates further studies of atomic interference 
and search of the novel trapping geometries. 
 
The aim of this article is to consider the 
influence of the reference frame rotations 
$\Omega$ on a superfluid confined by helical waveguide aligned 
along rotation axis $z$ \cite {Bhattacharya:2007,Okulov_helical:2012}. 
The interference of the counter propagating phase-conjugated Laguerre-Gaussian 
beams (LG) \cite{Woerdemann:2009} carrying orbital angular 
momentum (OAM) \cite {Allen:1992} 
gives the following rotating 
potential profile for the red-detuned alkali atoms (fig.1d): 
\begin{equation}
\label{TRAPOP1}
\ { U_{opt}{(z,r,\theta,t)}} \sim {\frac 
{U_0 \cdot r^{2|\ell|}}{(1+z^2/(k^2 {D_0}^4))} }
\exp \bigg{[}-{\frac {2r^2}{D_0^2(1+{ {z^2}/{k^2 D_0^4}})}}\bigg{]}
[1+\cos(2kz+2\ell\theta+\delta \omega t)],{\:}{\:}
\end{equation}
where the cylindrical coordinates $z,r,\theta$ are used, $D_0$ is 
the radius of LG \cite {Soskin:2008}, $\ell$ is 
vorticity, $k=2 \pi/ \lambda$ is wavenumber, $\delta \omega$ is 
angular Doppler shift 
\cite {Padgett:1998,Dholakia:2002} induced by rotation 
of the reference frame 
or emulated by rotation of Dove prism in a phase conjugated 
setup (fig.2) \cite {Okulov:2012josa}.
Transformation to the reference frame rotating synchronously 
with angular velocity $|\vec {\Omega}|=\Omega=\delta \omega /2 \ell$ 
with trapping helix leads to the time-dependent Gross-Pitaevskii equation 
(GPE) \cite {Pitaevskii:1999,Okulov_helical:2012,Berloff:2008}: 
\begin{equation}
\label{GPE_rot_frame}
{i \hbar}{\frac {\partial {\Psi}}
{\partial t}} = -{\frac {\hbar^2}{2 m}} 
\Delta {\Psi} +{ \tilde U_{opt}{(z,r,\theta)}}{\:}{\Psi} 
+{g}
|{\Psi}|{\:}^2 {\Psi}-\Omega \hat L_z {\Psi},
\end{equation}
where the stationary wavefunctions for the superfluid 
ensemble $\Psi=\Phi(z,r,\theta)\exp(-i\mu t/ \hbar)$ 
are given by: 
\begin{equation}
\label{GPE_static_in_rot_frame}
\mu {\Phi} = -{\frac {\hbar^2}{2 m}} 
\Delta {\Phi} +{\tilde  U_{opt}{(z,r,\theta)}}{\:}{\Phi} 
+{g}|{\Phi}|{\:}^2 {\Phi}+ 
\Omega {\:} i{\hbar}{\frac {\partial {\Phi}}{\partial \theta}},{\:}
\end{equation}
where ${\tilde  U_{opt}{(z,r,\theta)}}$ is time-independent 
optical potential. 
We evaluate linear and angular momenta of the superfluid 
ensemble in a helical 
container \cite {Okulov_helical:2012} and discuss 
the possibilities of rotations detection with this geometry. 
Hereafter the 
reference frame where laboratory is in rest will be called 
"laboratory frame" while reference frame collocated with 
viscous (classical) liquid completely trapped by rotating helical 
waveguide will be named "observer frame". 
Noteworthy the velocity 
vector $\vec V$ of "observer" with respect to "lab frame" 
has two components \cite {Okulov_plasma:2010}: the 
azymuthal velocity 
$\vec V_{\theta}= \vec {\Omega} \times \vec r$ stands for 
helix rotation around LG propagation axis $\vec z$ while helix 
pitch velocity $\vec V_{z}= ({\vec z}/z) {\delta \omega}/ 2 k$ 
is responsible for wavetrain translation along $\vec z$.  

\section{Twisted wavetrains in rotating frame}

In order to reveal the basic features of helical confinement 
let us decompose the potential in the time-dependent 
GPE \cite {Pitaevskii:1999,Okulov_helical:2012} taking into account 
the paraxiality of laser eigenmodes. Namely the spatial scales 
in descending order are $z_{R}>>D_0>>\lambda/2$, where 
$\lambda/2 \sim 5 - 0.5 \mu m$, $D_0 \cong 10 - 50 \mu m$ 
and $z_R \cong D_0^2 /{\lambda}$ (Rayleigh range). 
Thus the potential $\tilde U_{opt}$ can be expanded in a Taylor 
series by small parameter $z \lambda/(D_0^2)$ (inverse Fresnel number):
\begin{eqnarray}
\label{GPE_rot_frame}
{i \hbar}{\frac {\partial {\Psi}}
{\partial t}} = -{\frac {\hbar^2}{2 m}} [
\frac {\partial}{\partial z^2}+ \frac {\partial}{r {\:}\partial r}
r \frac {\partial}{\partial r}+\frac{\partial^2}{r^2 \partial \theta^2}]
\cdot {\Psi} +{g}
|{\Psi}|{\:}^2 {\Psi}-\Omega \hat L_z {\Psi}+
&& \nonumber \\
{\tilde U_{opt}(0)} \cdot {[1+\cos (2kz+2\ell\theta)]}\cdot
[{\frac {m {\omega_z}^2 z^2}{2 \cdot U_{opt}(0)}}+1] \cdot 
{\frac{r^{2|\ell|}}{D_0^{2|\ell|}}} 
\cdot exp[-{\frac{r^2}{D_0^2}}] \cdot{\Psi}.{\:}
\end{eqnarray}
This decomposition is valid due to relatively slow diffractive 
spread of helical waveguide 
during motion through LG-beam waist from $z=0$ point. Consider first 
the matter wavetrain $\Psi$ localized within Rayleigh range $z<<z_R$.  
The atomic waveguide is not expanded here, hence GPE is as follows: 
\begin{eqnarray}
\label{GPE_rot_frame_cylinder}
{i \hbar}{\frac {\partial {\Psi}}
{\partial t}} = -{\frac {\hbar^2}{2 m}} [
\frac {\partial}{\partial z^2}+ \frac {\partial}{r {\:}\partial r}
r \frac {\partial}{\partial r}+\frac{\partial^2}{r^2 \partial \theta^2}]
\cdot {\Psi} +{g}
|{\Psi}|{\:}^2 {\Psi}-\Omega \hat L_z {\Psi}+
&& \nonumber \\
{\tilde U_{opt}(0)} \cdot {[1+\cos (2kz+2\ell\theta)]}\cdot
{\frac{r^{2|\ell|}}{D_0^{2|\ell|}}} 
\cdot exp[-{\frac{r^2}{D_0^2}}] \cdot{\Psi}.{\:}
\end{eqnarray}
The variational anzatz  $\Psi_h (z,r,\theta,t)$ for the order parameter 
\cite {Malomed:2002} in a 
form of the superposition of the two phase-conjugated matter wave vortices 
\cite {Okulov_helical:2012} is written for helical trap within 
LG-beam waist ($|z|<<z_R$) as:

\begin{eqnarray}
\label{twisted_wavefunction}
\Psi_{_h} (z,r,\theta, t) =
\Psi_{\ell} (z,r,\theta, t)+ \Psi_{-\ell} (z,r,\theta, t) 
\cong {\:}{\:}{\:}{\:}{\:}{\:}{\:}{\:}{\:}{\:}{\:}{\:}{\:}
&& \nonumber \\
\Psi_{\pm \ell}(z=0){\cdot}({r}/{D_0})^{|\ell|}{\:}
\exp  {\:} \bigg{[} {\:} - {{r^2}/{{D_0}^2}} \bigg{]}\cdot
\exp  {\:} \bigg{[} {\:} - {{[z-z_{_1}(t)]^2}/{{Z_0}^2}} \bigg{]}\times 
{\:}{\:}{\:}{\:}{\:}{\:}{\:}{\:}{\:}
&& \nonumber \\
\bigg{\lbrace} 
{{\exp[ {-{{i \mu_{_f} t}/{\hbar}} + i k z+i\ell \theta} ]}}+
 {\exp[ {-{{i \mu_{_b} t}/{\hbar}} - i k z-i\ell \theta} ]}
 \bigg{\rbrace},{\:}{\:}
\end{eqnarray}
where $Z_0$ is a longitudinal size of the matter wavetrain, 
$\mu_{_f},\mu_{_b}$ are the chemical potentials of the 
"partial" wavefunctions $\Psi_{\pm \ell}$. 
The density $\rho$ of this atomic wavetrain has a helical shape 
$\rho (z,r,\theta,t)=|\Psi_h|^2$ and this helix rotates 
with angular velocity $\Omega=(\mu_{_f}-\mu_{_b})/(\hbar \cdot 2\ell )$. 
The qualitative analysis based on anzatz $\Psi_h$ shows that 
due to the absence of friction the center of mass of 
superfluid ensemble remains in rest in laboratory frame. 
But in "observer frame" the center of mass will move 
along straight line 
$z_1(t)=\pm V_{BEC}\cdot t=(\Omega \ell/k )\cdot t =(\delta \omega /2k )\cdot t$, 
where $\delta \omega /2k$ is the group velocity of the wavetrain (fig.3A).
The direction of this rectilinear motion is controlled by a product of 
the topological charge of trapping LG beams $\ell$ (the handedness of 
interference pattern) and projection of the vector 
of angular velocity $\vec {\Omega}$ 
of "observer frame" ${\Omega}$ on $z$ axis. 
Thus rotating observer will see 
$translational$ motion of the center of mass of twisted wavetrain 
in the absence of any externally applied force. Noteworthy the 
order parameter $\Psi_h$ is a solution of the time 
dependent GPE (eq.\ref{GPE_rot_frame_cylinder}) which describes purely 
condensed bosonic ensemble in the absence of thermal background. 

Outside the beam waist ($|z| \ge z_R$) the GPE with expanding trapping potential 
ought to be used (eq.\ref{GPE_rot_frame}).The variational anzatz 
for $|\Psi_h>$ is also a superposition \cite {Okulov_helical:2012}: 
 \begin{eqnarray}
\label{twisted_wavefunction_expanding}
\Psi_{_h} (z,r,\theta, t) =
\Psi_{\ell} (z,r,\theta, t)+ \Psi_{-\ell} (z,r,\theta, t) 
\cong {\:}{\:}{\:}{\:}{\:}{\:}{\:}{\:}{\:}{\:}{\:}{\:}{\:}
&& \nonumber \\
\Psi_{\pm \ell}(z=0){\cdot}{{({r}/{D_0})^{|\ell|}}}{\:}
\exp  {\:} \bigg{[} {\:} - 
{\frac {r^2}{{D_0}^2(1{+}iz/z_R)}} \bigg{]} \cdot
\exp  {\:} \bigg{[} {\:} - {{[z-z_{_2}(t)]^2}/{{Z_0}^2}} \bigg{]}\times 
{\:}{\:}{\:}{\:}{\:}{\:}{\:}{\:}{\:}
&& \nonumber \\
\bigg{\lbrace} 
{\exp[ {-{\frac {i \mu_{_f} t}{\hbar}} + i k z+i\ell \theta} ]}
+ {\exp[ {-{\frac {i \mu_{_b} t}{\hbar}} - i k z-i\ell \theta} ]}
\bigg{\rbrace}{\cdot}{{(1+iz/z_R)}}^{-1}{\:}{\:}{\:}.{\:}{\:}{\:} {\:}{\:}{\:}{\:}
\end{eqnarray}
Because red-detuned solutions are considered, the $|\Psi_h>$ is 
attracted by electrostatic potential $U_{opt}$ towards a beam 
waist $|z| \le z_R $ and wavetrain is reflected from low optical intensity 
regions $|z| > z_R$ towards the center of trap. Thus center of mass 
oscillates with effective equation of motion: 
\begin{equation}
\label{oscill_center_mass}
{\ddot z_{_2}(t)}+\gamma_Z \cdot {\dot z_{_2}(t)} +
[\omega_{Z}/(1+{2 \pi D_0/\lambda})]^2 
{z_{_2}(t)}=F_0 \cos (\Omega t){\:},
\end{equation}
where $\omega_{z} \cong \hbar / (m {z_R}^2)$ is a frequency of 
oscillations in a "longitudinal" well,
 $F_0$ is the averaged amplitude of "kicks", caused by reflections from 
trap boundaries located near $z \sim z_R$, $\gamma_{Z}$ is phenomenological 
damping constant.  The dissipation which is absent 
in initial GPE (eq.\ref{GPE_rot_frame}) 
appears in this model (eq.\ref{oscill_center_mass}) 
as phenomenological constant, arising because of leakage of atoms 
through potential barrier. 
The observer rotating with 
trap will see the onset of wavetrain oscillations (fig.3b) 
without any externally applied force. Consequently the helical 
waveguide for neutral atoms will 
transform rotation of lab frame into 
translational motion of the helical matter wave. 

\section{Angular momentum and translations of superfluid in 
rotating helical pipe}

It is instructive to evaluate exactly the macroscopic observables 
using variational wavefunction $|\Psi_h>$ 
(eqs.\ref{twisted_wavefunction},\ref{twisted_wavefunction_expanding}).
Because GPE (eqs.\ref{GPE_rot_frame},
\ref{GPE_rot_frame_cylinder}) gives coherent mean field wavefunction 
$\Psi$, the expectation values of linear $<\hat P_z>$ and angular momenta  
$<\hat L_z>$ may be obtained by averaging over macroscopic wavefunction 
\cite {Leggett:2001,Landau:1976}: 
\begin{equation}
\label{momentum2}
<P_z>_{h}=<{\Psi_{h}}|{ -i{\hbar}
{\frac {\partial }
{\partial z}}}|{\Psi_{h}}> =N {\hbar} (k_f - k_b),
{\:}{\:}{\:}{\:}{\:}{\:}
\end{equation}
taking into account the mutual orientation of the 
optical angular momenta of LG beams in a trap
\begin{equation}
\label{OAM1}
<L_z>_{h}=<{\Psi_{h}}|-i{\hbar}{\frac {\partial }
{\partial {\:} \theta}}|{\Psi_{h}}>=
N{\ell \hbar}(1\mp1), {\:}{\:}{\:}{\:}
{\:}{\:}{\:}
\end{equation}
where $N$ is the total number of bosons in a trap, the 
upper sign stands for the phase-conjugated LG-beams, i.e. for 
helical trap (fig.1d) \cite {Okulov:2008,Okulov_helical:2012}, 
while the bottom sign stands for 
sequence of toroidal traps aligned along the propagation axis $z$ (fig.1c).
As it easily seen from (eq.\ref{momentum2}) the ensemble moves along 
$z$ with momentum $\hbar (k_f - k_b)$ per particle  hence velocity 
of matter wave translation recorded by 
observer is $\Omega/(k_f + k_b)$. The angular momentum proves 
to be zero not only in $observer$ frame but in a $lab$ frame as well 
due to disappearance of the moment of inertia for rotating 
superfluid  \cite {Pitaevskii:1999}. 

There exists a remarkable difference  between proposed configuration 
(fig.2) and previously reported Sagnac 
matter wave interferometers \cite {Scully:1997} 
based on toroidal traps \cite {Phillips:2011},  
matter wave grating interferometers \cite {Pritchard:1997} and  
rotating low-dimensional traps where circulation of 
velocities around closed contour $\Gamma$ is 
essential \cite {Kartashov:2007,Sakaguchi:2007,Sakaguchi:2008}. 
The circulation is proportional to the 
angular momentum of a classical particle. 
In a simplest case of circular rotation in a plane 
perpendicular to $z$ axis the angular momentum equals to 
it's projection $L_z=m |\vec v| r$. The quantization of the 
angular momentum leads to discrete 
expectation values of operator $\hat L_z$ 
$<\Psi_n |-i \hbar \partial/\partial \theta |\Psi_n>=\pm n \hbar $ 
\cite {Landau:1976}. The quantum ensemble in a state $\Psi$ with 
zero circulation or zero angular momentum $n=0$ does not feel the 
rotation of container and may be referred to as an absolute reference 
frame. This is analog of the Meissner effect \cite {Leggett:2001} had 
been confirmed experimentally both for the 
liquid helium in annular \cite {Fairbank:1967} and 
for the trapped alkali gases in rotating toroidal  
traps. 

Nevertheless zero angular momentum state becomes  
thermodynamically unstable when the speed of 
angular motion $\Omega$ 
of container is increased above a certain 
critical value \cite {Pitaevskii:1999,Leggett:2001}. 
In this case the Feynman criterion for vortex appearance 
is applicable where critical velocity $V_{c_F}$ is 
due to Bohr-Sommerfeld quantization condition around 
vortex core giving rotational flows with quantized 
angular momentum $n \hbar$:
\begin{equation}
\label{Feynman criterion}
{\oint_{\Gamma} \vec v \cdot   d \vec r = \ell \hbar /m}, 
V_{c_F}= {n \frac { \hbar}{m R}} {\:}{\:}  ln (\frac {R}{R_c}).
\end{equation}
The values of critical velocity in this multiply connected geometry 
are smaller than those predicted by Landau 
criterion for superfluidity where critical velocity 
in homogeneous condensate is a tangent to 
dispersion curve near roton's 
minimum $V_{c_L}= {\epsilon(\vec p )/ |{\vec p}|}$ \cite{Feynman:1972}.
The same holds for modified Landau criterion proposed for the rotating 
helical flow \cite {Okulov_helical:2012} in the form 
$\Omega_{c}= D_0^{-1}{\epsilon(\vec p )/ |{\vec p}|}$ which correspond 
to excitation of rotons via centrifugal force and 
$|\vec V_{\theta}|=\Omega_{c} D_0$ is a tangential component of trapping 
helix velocity.

The angular momentum 
transfer from rotating environment to superfliud 
is facilitated by asymmetrical form of container \cite {Fairbank:1967} 
or externally 
imposed ponderomotive optical lattice 
potential  but the actual underlying 
mechanism includes interaction of quantum (superfliud) and classical 
(normal) components of the ensemble \cite {Pitaevskii:1999, Fetter:2009}.
This interaction leads to formation of unbounded vortices and 
vortex lattices \cite {Cornell:2006,Cornell:2007}. 
For the sufficiently fast rotations 
when $\Omega$ becomes comparable to the 
classical transverse oscillation frequency $\omega_{\bot}$ the 
lowest Landau level (LLL) state appears and this is 
accompanied by solitons formation \cite{Sakaguchi:2008} . 
Noteworthy the vortex lattices with 
also demonstrate a certain independence from external rotations imposed 
by revolving square optical lattices
\cite {Sakaguchi:2007}.

\section{Rotation of helical potential via angular Doppler effect}
	Experimentally the rotation of the optical interference pattern 
\cite {Okulov:2012josa,Cornell:2006} has a definite instrumental 
advantages over mechanical rotation \cite {Fairbank:1967}. The 
essential features of helical tweezer traps realized in 
\cite {Woerdemann:2009} are 
due to  apparent topological difference of toroidal 
and helicoidal interference patterns \cite {Bhattacharya:2007}(fig.2). 
The toroidal pattern appears when colliding photons have a parallel 
OAM's while helicoidal pattern 
arises for antiparallel photon's OAM's reflected from 
wavefront reversing mirrors (PCM) \cite{Okulov:2008}. 
In contrast to the conventional mirrors the PCM 
are essentially anisotropic optical elements 
\cite{Okulov:2008,Okulov:2008J,Okulov:2009,Okulov_plasma:2010}. 
Due to the internal helical 
structures, e.g. acoustical vortices in SBS mirrors, 
the PC mirror alters the angular momentum of 
the incident beam. Thus because the linear momentum $\vec P$ 
is altered 
too, the mutual orientation of $\vec P$ and $\vec L$ is conserved and 
the topological charge $\ell$ is not changed by an $ideal$ PC mirror. 
This may be also a $linear$ optical loop 
setup \cite{Okulov:2010josa} provided the $even$ number of reflections per 
a single loop roundtrip or the nonlinear optical PC retroreflector 
\cite{Okulov:2008,Woerdemann:2009}.

The alternation of the orbital angular momentum in 
PC mirror is accompanied by 
conservation identities \cite {Okulov:2012josa} 
for the frequencies 
of the incident and transmitted photons  and 
the projection of the angular momenta 
$L_z$ on $z-axis$. When Dove prism rotates 
with the angular velocity $\vec \Omega$ we have: 
\begin{eqnarray} 
\label {ang_moment_conserv_photon} 
I_{zz} \cdot \Omega + L_z =
I_{zz} \cdot {\Omega}^{'} + L_z^{'}
&& \nonumber \\
\hbar \omega_f + \frac {I_{zz}\Omega^2}{2}=
\hbar \omega^{'} + \frac {I_{zz}{\Omega^{'}}^2}{2},
\end{eqnarray}
where left hand sides of this system correspond to the 
incident photon and the right hand sides correspond to the 
transmitted one, $I_{zz}$ is the moment of inertia around $z-axis$. 
The difference of the angular velocities of the prism before and after 
the photon passage is due to reversal of 
OAM projection from $L_z=\pm \ell \hbar$ 
to $L_z^{'}=\mp \ell \hbar$ eq.(\ref{ang_moment_conserv_photon})  : 
\begin{equation}
\label {ang_velocity_change} 
\Omega - {\Omega}^{'} =- \frac {2 \ell \cdot \hbar}{I_{zz}}.
\end{equation}

From this follows that co-rotation increases the angular velocity 
of prism. Indeed the energy is transferred to the prism by 
the optical torque $|\vec T|=2 \ell \cdot P/{\omega_f}$, 
where $P$ is total power carried by LG 
\cite {Soskin:2008}. As a result Doppler frequency shift for the 
photon $\omega^{'}- \omega_f$ is negative in this case: 
\begin{equation}
\label {freq_shift_photon} 
\delta \omega = \omega^{'}- \omega_f = \frac {I_{zz}}{2 \hbar}
(\Omega-\Omega^{'})(\Omega^{'}+\Omega)=
-2 \ell \cdot \Omega - \frac{2\ell \cdot \hbar}{I_{zz}}.
\end{equation}

The else consequence of angular momentum conservation is that 
reversal of angular velocity of the prism $\vec \Omega$ 
alters the angular Doppler shift sign $\delta \omega$  and 
helical trapping potential 
${ \tilde U_{opt}{(z,r,\theta,t)}}$ rotates in opposite direction. The 
remarkable feature of this phase-conjugating laser 
interferometer technique 
\cite {Basov:1980, Okulov:2012josa} 
is that angular velocity of 
interference pattern  $\Omega=\delta \omega /2 \ell$ 
alters when vorticity of LG beam $\ell$ is altered (fig.2). 

\section{Conclusions}
The properties of superfluid ensemble in helical 
trap placed in rotating environment were analysed. 
For ultracold bosons ($10^{-6}\bf {K}$)  
trapped in vacuum by helical optical interference pattern the 
Gross-Pitaevskii mean field theory was applied. The variational 
approach indicates the possibility 
of detection of the slow rotations of observer by means of measurement 
of mechanical momentum of atomic cloud. Namely 
the rotating observer will detect the appearence of the 
translational rectilinear 
motion and oscillations of matter wavetrain near 
equilibrium position of the trap without any externally applied 
force. 
The orbital angular momentum of trapping beams 
\cite {Padgett:1998, Barnett:2002} and rotational 
Doppler shift \cite {Dholakia:2002, Okulov:2012josa} 
are the key components for the proposed experiment. 

Hopefully the GPE-based  model considered above 
is qualitavely applicable to the flow of 
liquid $^4He $ at 
$\bf T_{\lambda} {<2.17}^o \bf {K}$ temperatures through a 
$\mu m$-size capillary \cite {Kapitza:1938,Allen:1938,Kapitza:1941}.
In particular the $\mu m$-size helical pipe filled by 
$^4He $ superfluid and placed to rotating container 
is expected to eject quantum fluid outwards the pipe 
with translational speed $V_z=\Omega \ell /k$ defined by angular frequency of 
container $\Omega$, the pitch of the helix $\lambda=k/2 \pi$ and 
winding number $\ell$. The handedness of the helical 
pipe $\pm \ell$ 
defines the direction of superfluid translation along 
positive or negative direction of $\vec z$ axis. Apparently the 
visibility of effect should be much better for the temperatures much below 
$\lambda$-point because of higher density of the condensate 
fraction $\rho_s \rightarrow \rho $.

Qualitative picture 
is expected to be exactly opposite for classical fluid. 
In a frame collocated with rotating observer 
the classical fluid will stay in rest because of 
nonzero viscosity. In a rest (laboratory) frame the classical fluid 
will be pushed lengthways $z$-axis by rotating helical pipe. Thus the 
classical flow is controlled dynamically, by Navier-Stockes 
equation, i.e. by Newtonian forces and viscosity. On the 
contrary the quantum flow in rotating helical pipe is due 
to the purely kinematical reasons. In both cases the direction  
of translational velocity is defined 
by a sign of $\ell \cdot \Omega$ product.

\begin{figure}
\center{ \includegraphics[width=12cm]{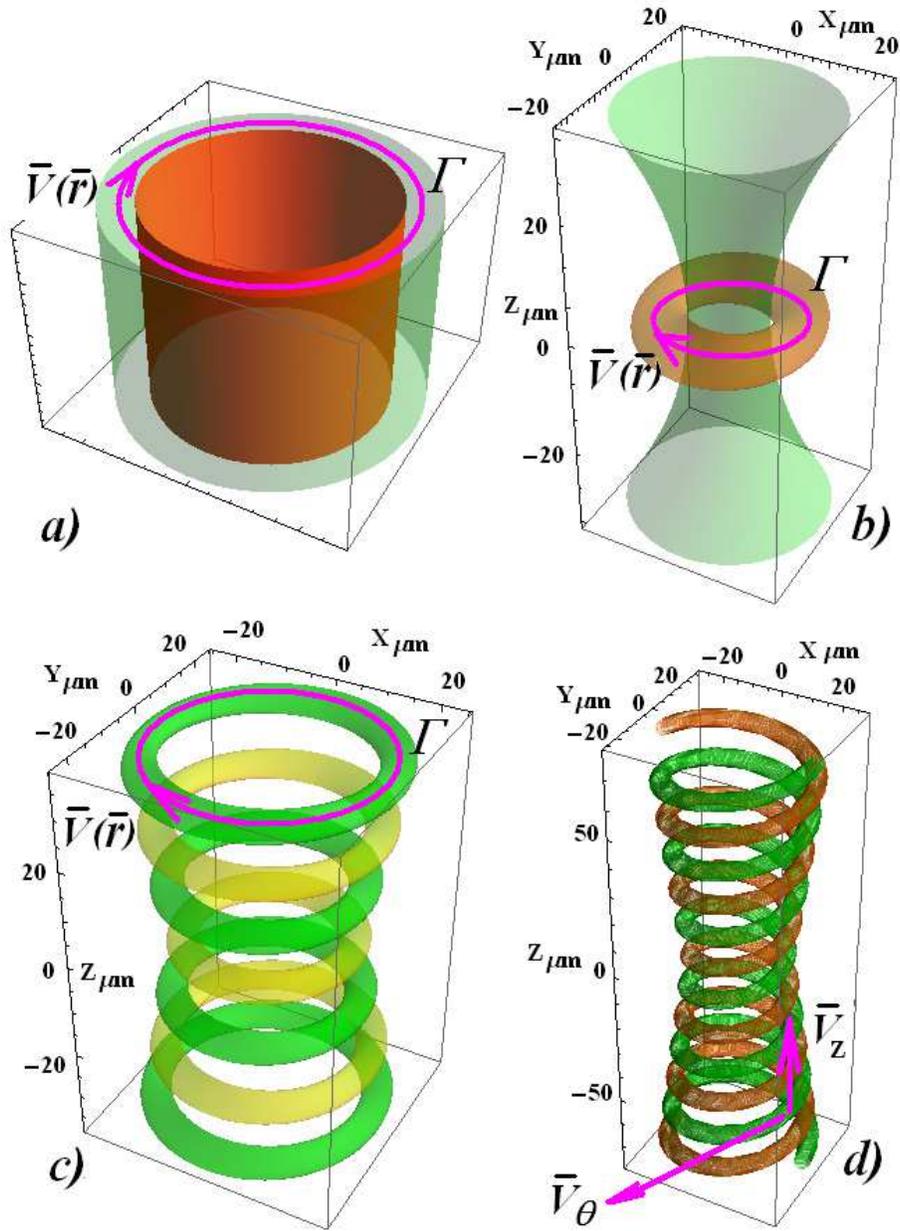}}
\caption{Multiply connected potentials, $\vec v (\vec r)$ 
is local velocity vector, $\Gamma$ - trajectory (contour) for 
calculating circulation : 
a) Annular cylinder 
in Hess-Fairbank angular momentum experiment 
\cite {Fairbank:1967}, b) 
Toroidal potential in LG beam waist \cite {Phillips:2011,Konotop:2007} 
and in magnetic 
trap pierced by repulsive laser 
\cite {Phillips:2011}, 
c) Sequence of toroidal 
wells separated by $\lambda/2$ produced by two confocal $LG_{1,0}$ beams 
(with parallel OAM's) \cite{Rempe:2007,Okulov:2008,Woerdemann:2009}. 
d) Helical interference pattern of the 
two phase-congugated $LG_{1,0}$ beams (with opposite OAM's)
\cite{Rempe:2007,Okulov:2008,Woerdemann:2009,Okulov_helical:2012} and 
velocity of observer $\vec V(\vec r)= \vec V_z + \vec V_{\theta}$, which 
moves together with trapped classical liquid.} 
\label{fig.1}
\end{figure} 
 
\begin{figure}
\center{ \includegraphics[width=12cm]{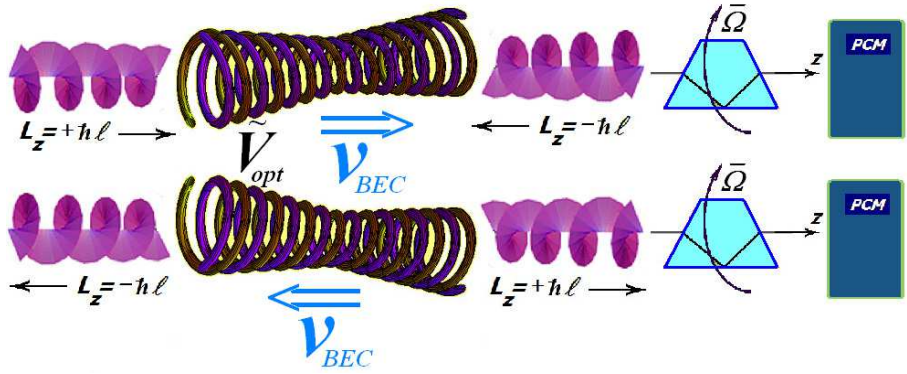}}
\caption{Helical optical trapping potential $\tilde U_{opt}$ formed 
by two phase-conjugated  
optical vortices with opposite orbital angular 
momenta $\pm \hbar \ell$. Two mutual orientations 
of the angular momenta of vortices give the interference patterns 
of right handedness ($+\ell=1$ upper) and of left 
handedness ($-\ell=1$ bottom) respectively. The  
rotation of interference pattern with angular frequency ${\Omega} $ 
is induced by 
rotating Dove prism located between trapping volume and 
phase-conjugating mirror PCM. This rotation 
is equivalent to rotation of the reference frame. The twisted 
BEC wavetrain in 
rotating helical atomic waveguide moves with a speed $V_{BEC}=\Omega \ell /  k$ 
in a positive or a negative $z$ direction 
accordingly to the sign of $\Omega \ell $.}
\label{fig.2}
\end{figure} 
 
\begin{figure}
\center{  \includegraphics[width=12cm]{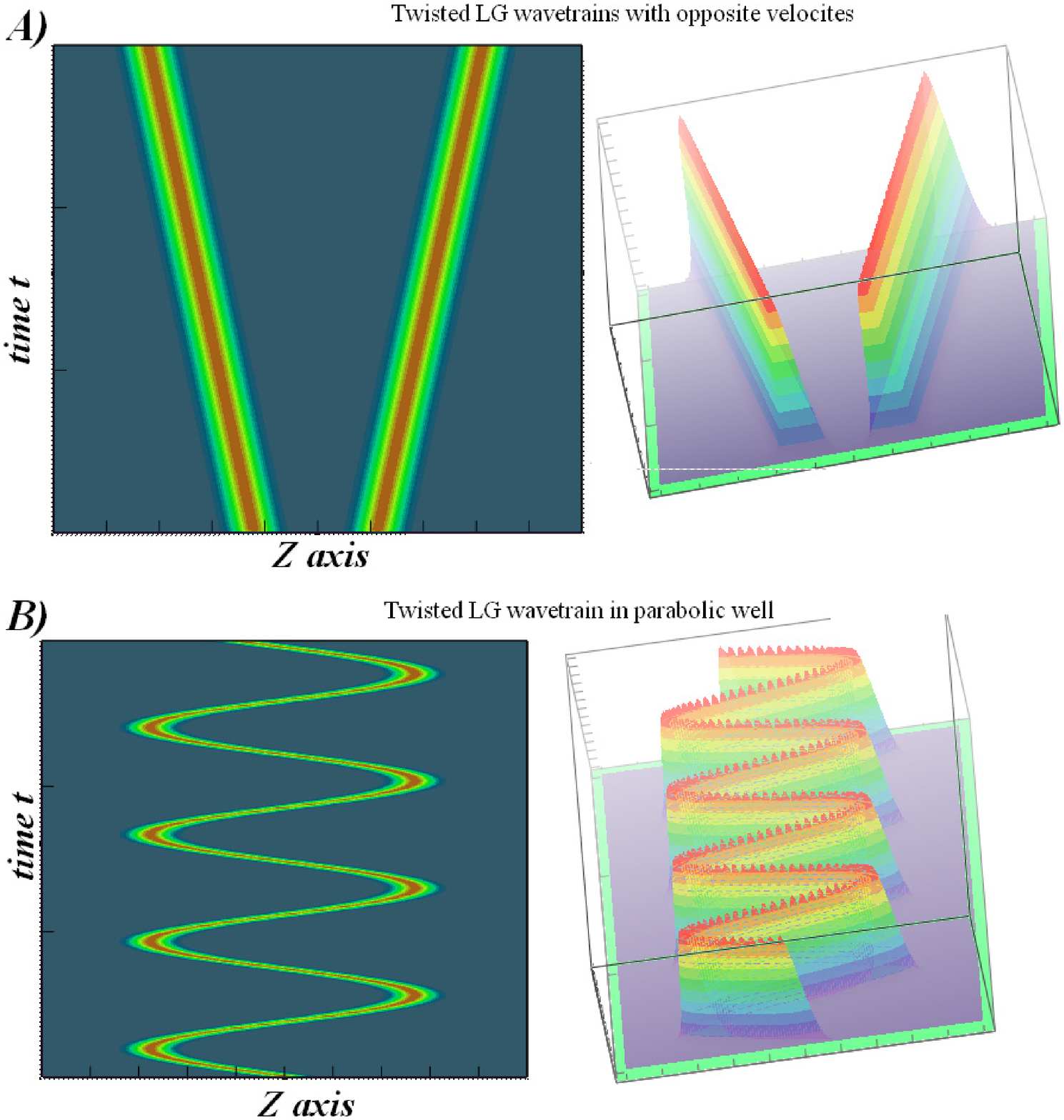}}
\caption{Twisted matter wavetrain with Gaussian $z$-envelope in 
rotating helical atomic waveguide.
a) In a waveguide with homegeneous ($z$-independent) width of
potential $U_{opt}$
the wavetrain will move with a constant speed 
$V_{BEC}=\pm \Omega \ell / k$ 
either positive or negative direction of $z$. 
b) In a waveguide with parabolic $z$-dependent potential $U_{opt}$ 
having width $D_0=D_0(1+z^2/z_R^2)$) 
the wavetrain will experience oscillations between 
reflection points \cite {Konotop:2007}.}
\label{fig.3}
\end{figure}


\begin{thebibliography}{99}

\bibitem{Fairbank:1967} 
{G. Hess and W. Fairbank,}{Phys.Rev.Lett.}, {\bf 19}, 216 (1967).

\bibitem{Feynman:1972}
{R.P.Feynman,}{\itshape "Statistical mechanics"}, 
Ch.11,(1972) Reading, Massachusetts. 

\bibitem{Phillips:2011}
{A. Ramanathan, K. C. Wright, S. R. Muniz, M. Zelan, 
W. T. Hill, III, C. J. Lobb, K. Helmerson, W. D. Phillips, 
and G. K. Campbell,} {Phys.Rev.Lett.} {\bf 106}, 130401 (2011). 

\bibitem{Wright:2001}
{ E. M. Wright, J. Arlt, and K. Dholakia,}
{Phys.Rev.A} {\bf 63}, 013608  (2002).

\bibitem{Pitaevskii:1999}
{F. Dalfovo, S.Giorgini, S.Stringari, L.P.Pitaevskii},
{Rev.Mod.Phys.}{\bf 71},463(1999).

\bibitem{Rempe:2007}
{T. Puppe, I. Schuster, A. Grothe, A. Kubanek, K. Murr, P.W.H. Pinkse, 
and G. Rempe,} {Phys.Rev.Lett.}, {\bf 99}, 013002 (2007).

\bibitem{Okulov:2008}
{A.Yu.Okulov,} J.Phys.B., {\bf 41},101001 (2008).

\bibitem{Fetter:2009} 
{A.L.Fetter}, {Rev.Mod.Phys.} {\bf 81}, 647 (2009).

\bibitem{Leggett:2001} 
{A.J. Leggett}, {Rev.Mod.Phys.} {\bf 73}, 307-356 (2001).
 
\bibitem{Konotop:2007}
{Yu. V. Bludov, and V.V.Konotop, }
{Phys.Rev.A} {\bf 75}, 053614 (2007).

\bibitem{Scully:1997}
{M.O.Scully, M.S.Zubairy},{\itshape "Quantum optics"},
Ch.4, (Cambridge University Press) (1997).

\bibitem{Okulov:2010josa}{A.Yu.Okulov,} J. Opt. Soc. Am. B 
{\bf 27}, 2424-2427 (2010).
 
\bibitem{Bhattacharya:2007} {M.Bhattacharya,} 
{Opt.Commun.} {\bf 279}, 219 (2007).

\bibitem{Okulov_helical:2012} {A.Yu.Okulov,} {Phys.Lett.A}, {\bf 376}, 
650-655 (2012).

\bibitem{Woerdemann:2009}
{M.Woerdemann, C.Alpmann and C.Denz,}
{Opt. Express}, {\bf 17}, 22791(2009). 

\bibitem{Allen:1992}{L.Allen, M.W.Beijersbergen, R.J.C.Spreeuw 
and J.P.Woerdman, }{Phys.Rev.A}, {\bf 45}, 8185-8189 (1992). 
 
\bibitem {Soskin:2008}{A. Bekshaev, M.Soskin and M. Vasnetsov},
{\itshape "Paraxial Light Beams with Angular Momentum",}
{Nova Science}(2008).

\bibitem{Padgett:1998}{Courtial J., Robertson D. A., Dholakia K., Allen L., 
Padgett M. J.,} {Phys.Rev.Lett.}, {\bf 81},4828-4830(1998).

\bibitem {Dholakia:2002}J. Arlt, M. MacDonald, L. Paterson, 
W. Sibbett,K. Volke-Sepulveda and K. Dholakia, 
Opt. Express, {\bf 10} , 844 (2002).

\bibitem{Okulov:2012josa}{A.Yu.Okulov,} J. Opt. Soc. Am. B 
{\bf 29}, 714-718 (2012).

\bibitem{Berloff:2008} {J.Keeling and N. G. Berloff},
{Phys.Rev.Lett.,} {\bf 100},250401 (2008).
 
\bibitem{Okulov_plasma:2010}{A.Yu.Okulov,} 
{Phys.Lett.A}, {\bf 374},4523-4527 (2010).
 
\bibitem{Malomed:2002}{B.A.Malomed}, 
{\itshape "Variational methods in nonlinear fiber optics and related fields"},
Progress in Optics,(E.Wolf, Editor: North Holland, Amsterdam ) 
{\bf 43},  69-191 (2002).

\bibitem{Landau:1976} 
{L.D. Landau and E.M. Lifshitz},\textit 
{"Quantum Mechanics"},{ Butterworth-Heinemann, Oxford}(1976).

\bibitem{Pritchard:1997}
{A. Lenef, T.D. Hammond, E. T. Smith, M.S.Chapman, R.A.Rubenstein, and 
D.E.Pritchard,}{Phys.Rev.Lett.}, {\bf 78},  760-763 (1997).

\bibitem{Kartashov:2007}{Y.V.Kartashov, B.A. Malomed, and L.Torner}, 
{Phys.Rev.A} {\bf 75}, 061602 (2007).

\bibitem{Sakaguchi:2007}{H.Sakaguchi and B.A. Malomed}, 
{Phys.Rev.A} {\bf 75}, 013609 (2007).

\bibitem{Sakaguchi:2008}{H.Sakaguchi and B.A. Malomed}, 
{Phys.Rev.A} {\bf 78}78, 063606 (2008).

\bibitem{Cornell:2006} S.Tung, V.Schweikhard, and E.A.Cornell,
{Phys.Rev.Lett.,} {\bf 97},240402 (2006).

\bibitem{Cornell:2007} S.Tung, V.Schweikhard, and E.A.Cornell, 
{Phys.Rev.Lett.,} {\bf 99}, 030401 (2007).
 
\bibitem{Okulov:2008J}{A.Yu.Okulov,} JETP Lett., {\bf 88}, 631 (2008).

\bibitem{Okulov:2009} {A.Yu.Okulov,} {Phys.Rev.A }, {\bf 80}, 013837 (2009).

\bibitem{Basov:1980}{N.G.Basov, I.G.Zubarev, 
A.B.Mironov, S.I.Mikhailov and A.Y.Okulov}, JETP, {\bf 52}, 847 (1980).

\bibitem{Barnett:2002}
{J.Leach,M.J.Padgett,S.M.Barnett,S.Franke-Arnold, and J.Courtial,}
{Phys.Rev.Lett.}, {\bf 88}, 257901(2002).

\bibitem{Kapitza:1938}{P.L.Kapitza,} Nature, {\bf 141}, 74 (1938).

\bibitem{Allen:1938}{J.F.Allen, A.D.Misener,} Nature, {\bf 141}, 74 (1938).

\bibitem{Kapitza:1941} {P.L.Kapitza,} Phys.Rev., {\bf 60}, 354 (1941).

\end{thebibliography}
\end{document}